\documentclass[10pt, a4paper]{article}

\usepackage{graphicx}
\usepackage{subcaption}
\usepackage{authblk}
\usepackage{amsmath}
\usepackage[english]{babel}
\usepackage{array}

\usepackage{cite}
\usepackage{siunitx}
\begin{document}


\title{Analyzing Mass Media influence using natural language processing and time series analysis}

\author[1]{Federico Albanese \thanks{falbanese@dc.uba.ar}}
\author[2,3]{Sebasti\'an Pinto}
\author[4]{Viktoriya Semeshenko}
\author[2,3]{Pablo Balenzuela}

\affil[1]{Instituto de Investigaci\'on en Ciencias de la Computaci\'on, CONICET- Universidad de Buenos Aires, Argentina}
\affil[2]{Departamento de F\'isica, Facultad de Ciencias Exactas y Naturales, Universidad de Buenos Aires, Av. Cantilo s/n, Pabell\'on 1, Ciudad Universitaria, 1428, Buenos Aires, Argentina.}
\affil[3]{Instituto de F\'isica de Buenos Aires (IFIBA), CONICET, Av. Cantilo s/n, Pabell\'on 1, Ciudad Universitaria, 1428, Buenos Aires, Argentina.}
\affil[4]{Instituto Interdisciplinario de Econom\'ia Pol\'itica (IIEP-BAIRES), Facultad de Ciencias Econ\'omicas, CONICET-UBA, Argentina}

\date{December 2019}
             
\maketitle

\begin{abstract}
\par A key question of collective social behavior is related to the influence of Mass Media on public opinion. Different approaches have been developed to address quantitatively this issue, ranging  from field experiments to mathematical models. In this work we propose a combination of tools involving natural language processing and time series analysis. We compare selected features of mass media news articles with measurable manifestation of public opinion. We apply our analysis to news articles belonging to the 2016 U.S. presidential campaign. We compare  variations in polls (as a proxy of public opinion) with changes in the connotation of the news (sentiment) or in the agenda (topics) of a selected group of media outlets. Our results suggest that the sentiment content by itself is not enough to understand the differences in polls, but the combination of topics coverage and sentiment content  provides an useful insight of the context in which public opinion varies. The methodology employed in this work is far general and can be easily extended to other topics of interest.

\end{abstract}


\section{Introduction}

\par Mass Media play one of the important roles in the process of public opinion formation. Beyond informing about facts and events, mass media give an interpretation about such events, providing individuals a way to understand their relevance. Through its capacity to reflect reality from its own perspective, media  determine the relative importance given to different topics, a process known as agenda-setting. The agenda-setting theory is usually summarized in the quote ``maybe media does not tell you what to think, but what to think about" \cite{mccombs1972agenda, mccombs2005look}. 
In other words, media can tell you what is and what is not important, and to what extent. 
Therefore, the agenda-setting power of mass media produce an important effect, that acquires relevance, for instance, in the political opinion formation  during electoral contexts \cite{fortunato2016intersection}.

\par Previous research has shown how public perception of a political event is modified by mass media  \cite{besley2002political}. 
The influence is usually manifested on the basis of topics emphasized and omitted by the media \cite{brians1996campaign}. 
In addition, other studies demonstrated that reading different media in a sustained manner leads individuals to modify their political ideology, aligning their votes with the editorial viewpoint of a given newspaper \cite{oberholzer2009media}.
Gerber and Dean \cite{gerber2009does} studied the 2005 governor elections in the state of Virginia (US), and observed that the newspaper's reading affected the decisions of a finite number of voters, and produced induced changes in the perception of politicians.

\par In the last decades, given the availability of data and computational resources, the quantitative analysis of mass media influence has been addressed from different perspectives.

\par On one hand, the emergence of social media and availability of online data enriched the research about  mass media impact. For instance, King et al. \cite{king2017news} detected an increase in the number of tweets about a specific topic  after being exposed to related news. 
Yasseri and Bright \cite{yasseri2016wikipedia} showed that consulting for the number of mentions of a candidate in Wikipedia produces changes in vote shares for particular parties, regardless of whether those were positive or negative. 
The spreading and consumption of fake news in social media was analyzed in \cite{allcott2017social}, where the authors show that people tend to share fake news that reinforce their ideological bias.
The role of the news' sentiment (positive or negative connotation) has been explored within the framework of how the connotation is related to the variation of economic indicators \cite{soroka2015s}, how it affects public expectation about economy \cite{lischka2015follows, hopkins2017does}, or how it shapes the public opinion about a given prominent issue \cite{de2006media}. In the same line, \cite{bae2012sentiment, gorodnichenko2018social} addressed how the sentiment of bots' or influential users' tweets induces the connotation of their followers expressions in Twitter.
Other techniques such as topic modeling were also employed in order to describe topics dynamics in media \cite{koltsova2013mapping, korencic2015getting, Pinto2019}, and how it is related to audience response.

\par On the other hand, computational models have also been implemented in order to evaluate different mechanisms of interactions between mass media and individuals, and how mass media influence  the formation of collective public opinions  \cite{shibanai2001effects, gonzalez2006local, pinto2016setting}. In this kind of models, the behavior of mass media is generally not grounded on data, but the incorporation of relevant information contained in news articles, as is analyzed in this work, gives a powerful tool to explore different scenarios of interactions between mass media and public opinion \cite{Albanese2019}.
\par In this paper, we  study the relationship between mass media and public opinion  using a combination of sentiment analysis and topic detection of news articles. As a proof of concept, we  apply this methodology to a particular case study which is the 2016 US presidential campaign, analyzing news articles where the involved candidates are mentioned. We consider the number of mentions of each candidate, their sentiment content  and the evolution of the relative coverage of a set of topics of the same articles (political media agenda). Beyond the particular case studied, the methodology introduced in this work is far general and can be extended easily to other case of interest.

\par This work is organized as follows: In section \ref{sec:matmeth} we describe  the natural language processing tools applied to the news articles. In section \ref{sec:data} we describe the  data of the studied case. In section \ref{sec:results} we analyze the time series of the polls,  the number of mentions of the candidates, the sentiment analysis of the news content, the topic evolution of the political media agenda, and the sentiment analysis discriminated by topic. We measure the Spearman correlation and the Granger causality to draw useful conclusions. Finally, we discuss these results in section \ref{sec:discussion}.

\section{Data: The 2016 US Presidential Elections}
\label{sec:data}

\par In 2016, a presidential election took place in the United States, facing up the Democrat candidate Hillary Clinton and the Republican candidate Donald Trump, who finally won the election.
We centered our analysis in two types of data: Polls, as a proxy of public opinion  and news articles from four of the main US mass media.  The whole analyzed period comprises from $28^{th}$ of July  (last  party convention, where the candidates were formally defined) to the $8^{th}$ of November of 2016 (election date).

\subsection{Polls data}
\par We analyzed a total of 263 national surveys conducted by different agencies (an average of 2.7 surveys per day), in which the forecast of the votes of each candidate was measured in a gap of a few days (around 3-5 days). This data, which shows the result of different pollsters, was downloaded from the Real Clear Politics website \cite{RCP}. All national surveys presented in this work belong to a demographic balanced sample. Polls data fulfill the terms and conditions specified in the site Real Clear Politics and are available as supporting material.

\par
\begin{figure}[h]
\centering
\includegraphics[width = \textwidth]{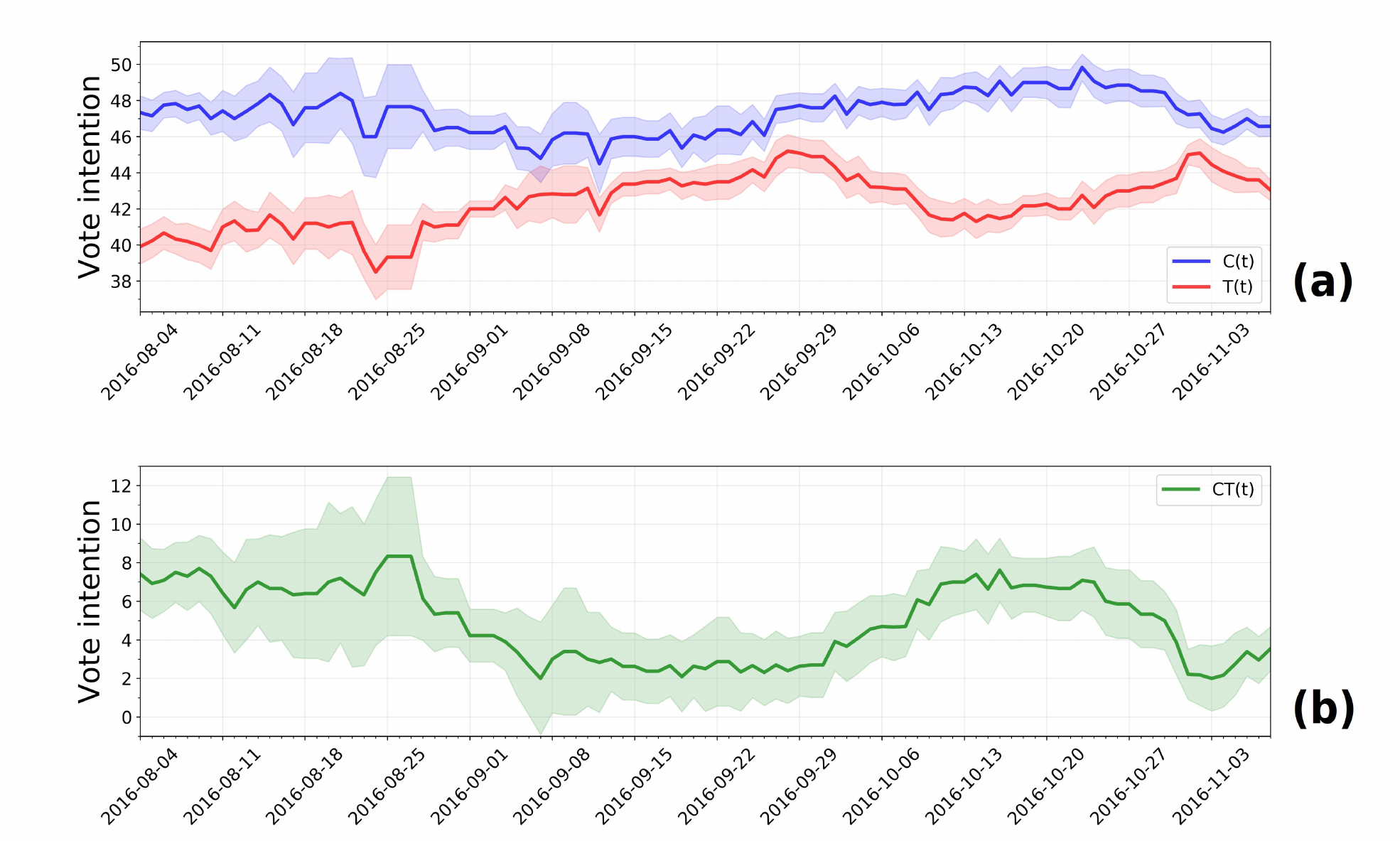}
\caption{\textbf{Survey curves of 2016 US presidential elections}. 
Top: The curves of Hillary Clinton (Democrat candidate) and Donald Trump (Republican candidate) for the 2016 US presidential election. Bottom:  The spread between these curves.}
\label{fig1}
\end{figure}

 Figure \ref{fig1} displays the time series of potential percentage of votes for each candidate (top panel), and the difference between these time series (the percentage of Clinton minus the percentage of Trump, bottom panel). Each point in the time series represent the average of the preceding week. In other words, a 7-days sliding window average was used, which means that each point in the time series takes into account an average of 19 polls.

\par Figure \ref{fig1} shows that Clinton kept up an advantage over Trump during all the time period. However, this advantage was affected by drops in the middle and close to the end of the period. By looking at the individual time series, we can see that the initial decreasing of the advantage is due to the gradual ascending of the Trump's intention to vote until October. After that, his percentage went down sharply until it was recovered near the election date, which explains the last decrease in the Clinton's advantage, which slightly increased in the last days.

\subsection{Mass Media data}

\par  We selected the online editions of The New York Times, Fox News, CNN and USA Today to perform our analysis. The selection criteria  is that they are the most popular mass media outlets in term of online searches in all U.S. territory (see Appendix A).  For instance, the first one is a classical newspaper based in New York City with worldwide influence and readership and the second and third one are cable television news channels which broadcast to many countries all around the world. USA Today is an internationally distributed American daily, middle-market newspaper.

\par  We selected the articles which contain at least the name of one of the two main candidates: Hillary Clinton (Democrat) and Donald Trump (Republican).
The analyzed corpus is made up of a total of $15175$ articles: $5672$ from The New York Times, $5750$ from Fox News, $2920$ from CNN, and $833$ from USA Today.
We include all articles that mention one or both candidates irrespective of the section they belonged to. The full corpus is available as supporting material.

\section{Methods}
\label{sec:matmeth}

\subsection{Text Mining Methods}
\label{sec:textmining}

\par We focused our analysis in text mining techniques, from which we extract useful information by applying both, sentiment analysis and topic detection,  to news articles whose corpus will be detailed bellow.

\subsubsection{Sentiment Analysis}
\label{subsec:sentimentanalysis}

\par In order to measure the frequency of positive and negative mentions for a given candidate, we implemented a sentiment analysis algorithm.
The sentiment analysis was performed through deep recursive models for the semantic composition applied to sentiment trees \cite{socher2013recursive}, in particular by the Stanford CoreNLP implementation of natural language processing \cite{manning2014stanford}. 
This algorithm consists of assembling a tree from the grammatical structure and a syntactic analysis of each phrase. Then, each word (node) is assigned a sentiment value, taken from a database: very positive, positive, neutral, negative or very negative. 
In addition, this algorithm takes into account if the words are intensifiers, appeasers, deniers, etc. The algorithm assigns a sentiment value  to each node starting from the inner nodes.  After several iterations, it ends up assigning the corresponding sentiment value to the total phrase. 

\par There exist several algorithms to perform sentiment analysis, such as those based on the extraction of characteristics of the sentences \cite{doddi2014sentiment} or lexicon-based approaches to opinion mining \cite{taboada2011lexicon, muhammad2016contextual}. Given that our corpus of news is formed by grammatically correct sentences,  the Stanford CoreNLP is the proper algorithm to perform sentiment analysis.

\subsubsection{Topic Detection}
\label{sec:topicdetection}

\par In addition to the sentiment analysis, we also performed a topic detection on the corpus of the news articles using unsupervised learning techniques as was implemented in \cite{Pinto2019}.

\par We represent news articles as numerical vectors through the \emph{term frequency - inverse document frequency (tf-idf)} representation. The value of each component is given by the frequency of each term in the text (tf) weighted by a measure of specificity (idf) \cite{xu2003document}. 
Vectors are then compiled in a document-term matrix \emph{M}, of dimension $d$ (number of documents) per $t$ (number of terms, $t=56979$ in our work), which is given by the total amount of words in the corpus after stop-word removal \cite{Amaral2019}.  
On the other hand, we detect the main topics in the corpus by performing \emph{non-negative matrix factorization (NMF)} \cite{xu2003document, lee1999learning} on the document-term matrix (\emph{M}). A topic is defined as a group of similar articles which roughly talk about the same subject. 
Let us mention, that analogous results were obtained by applying Latent Dirichlet Allocation (LDA) method \cite{blei2003latent} on the same corpus.
\par NMF decomposes matrix \emph{M} as the product of two non-negative matrices, $H$ and $W$ (see equation Eq.(\ref{eq:nmf})), where the first one is a document-topic matrix 
and gives us the representation of the documents in the space of topics, while the second one is a topic-term matrix and brings the topics described in the space of terms, from where we can extract the keywords which define each one. The inner dimension $n$ in Eq.(\ref{eq:nmf}) is the number of expected topics, which is a parameter that must be set before the decomposition.
\begin{equation}
    M^{(d \times t)} \sim H^{(d \times n)} \cdot W^{(n \times t)}
    \label{eq:nmf}
\end{equation}

\par In order to calculate the coverage of mass media, we estimate the amount of articles and their relative importance in each topic. For this sake we define the weight of the topic $i$ ($T_i$) as the product of the amount of news articles (weighted by their degree of membership) and the length of the article. The coverage can be defined on a daily basis (time-dependent distribution) or for the whole period (average distribution). Eq.(\ref{eq:topic_weight}) shows the coverage for a single day $d$, 

\begin{eqnarray}
\label{eq:topic_weight}
T_i(d) = \sum_j l(j) \cdot h_{ji} \cdot \delta_{d_j,d},
\end{eqnarray}

where $l(j)$ is the number of words in the document $j$; $h_{ji}$ (element of matrix \emph{H}) is the degree of membership of document $j$ on topic $i$; $d_j$ is the date of document $j$; and $\delta$ is the Kronecker delta.  Given that each document vector can have all non-zero components, it is allowed that a document contributes to more than one topic weights.
In order to reduce noise, we finally apply a linear filter with a seven day wide sliding window, and normalize the temporal profiles. 

\subsection{Correlation and causality measures}

\subsubsection{Spearman Correlation}

\par Spearman correlation coefficient is a non-parametric measure of rank correlation. It assesses how well the relationship between two variables can be described using a monotonic function. While Pearson correlation measures linear relation between the two variables, Spearman correlation  assesses the monotonic relationships between them. Before computing the Spearman correlation, we removed the respective linear trend, if there was any, from all time series in order to avoid spurious correlations.

\subsubsection{Granger Causality}
\label{sec:Granger}

\par The Granger causality test \cite{granger1969investigating} determines if one time series is able to forecast another one. 
Given a stationary time series, $x_t$, modeled by an auto-regressive-moving-average model Eq.(\ref{eq:Granger1}), the Granger casuality test basically determines if the model of Eq.(\ref{eq:Granger2}) is better than the model of Eq.(\ref{eq:Granger1}). In other words, this means that the additional information provided by a second time series, $y_{t-\tau}$, improves the forecasting of $x_t$.

\par The amount of terms that must be included in Eq.(\ref{eq:Granger1}) and (\ref{eq:Granger2}) are determined by studying both the autocorrelation and the partial correlation of $x_t$, where $w_t$'s are white noise terms, and $\theta$'s, $\phi$'s, and $\beta$ are just coefficients \cite{shumway2017time}. If $\beta$ is significantly different from zero, we may say that $y_t$ has a causal relation with $x_t$. Notice that when $\beta$ is zero, the model of Eq.(\ref{eq:Granger1}) is recovered.

\begin{equation}
    x_t = \sum_i \phi_i x_{t-i} + \sum_j \theta_j w_{t-j} + w_t 
    \label{eq:Granger1}
\end{equation}
\begin{equation}
    x_t = \sum_i \phi_i x_{t-i} + \sum_j \theta_j w_{t-j} + w_t + \beta y_{t-\tau}
    \label{eq:Granger2}
\end{equation}

\section{Results}
\label{sec:results}

Here we analyze the polls data and the news articles (by extracting the sentiment content and the topic decomposition) in the electoral period from the $28^{th}$ of July to the $8^{th}$ of November of 2016. This period comprehends since the last 2016 party convention, where the candidates were formally defined, until the election date.

\subsection{Total number of mentions}
\label{sec:mentions}

\begin{figure}[h]
\centering
\includegraphics[width = \textwidth]{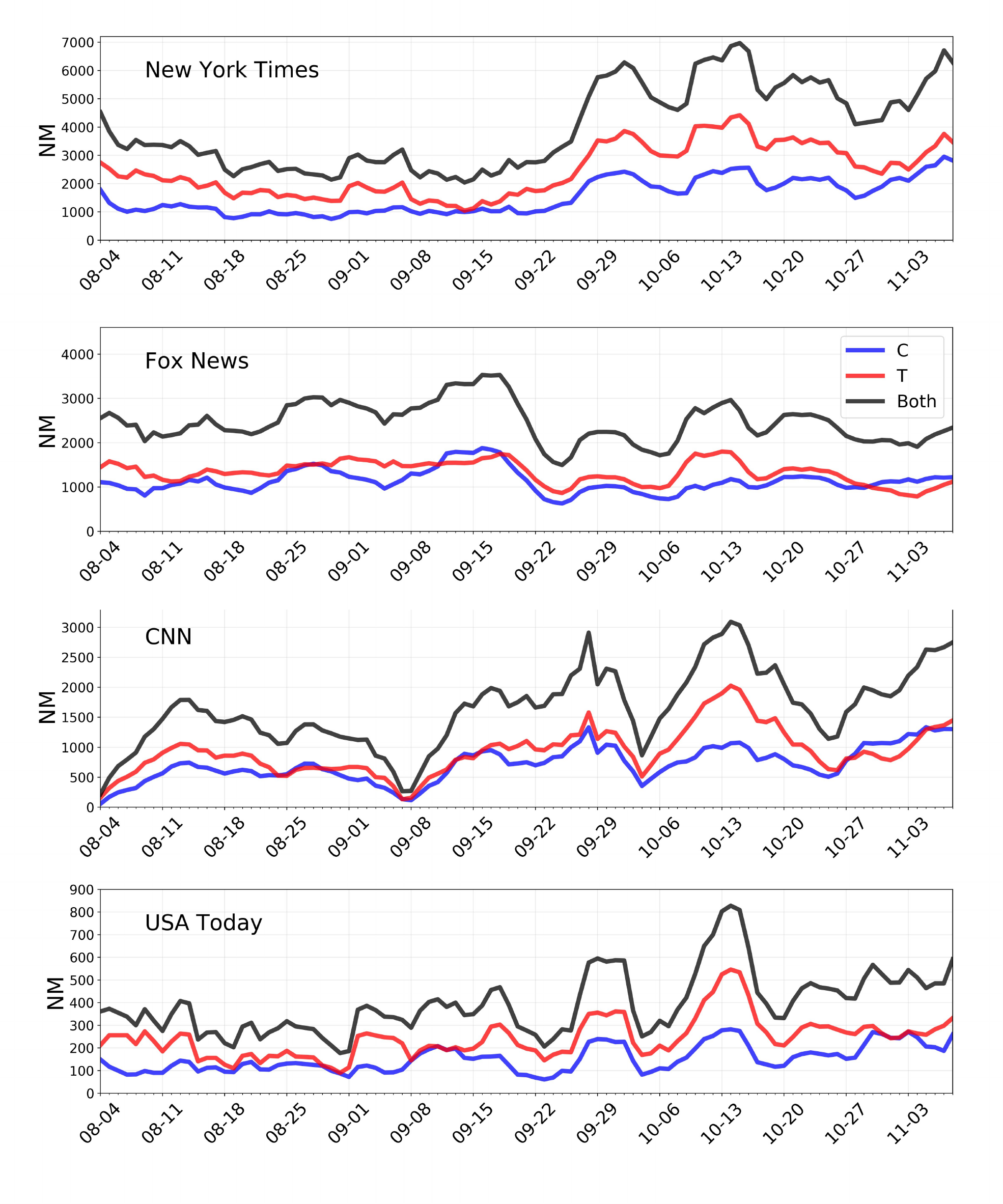}
\caption{\textbf{Time series of the total number of mentions}.
The curves shows the total number of mentions of Hillary Clinton (in Blue), Donald Trump (in red) and the sum of both of them (in black) for each media outlet (from top to bottom, New York Times, Fox News, CNN, and USA Today).}
\label{fig2}
\end{figure}

As a first approach, we  compared the time series of the surveys with the total number of mentions of each candidate in both media \cite{yasseri2016wikipedia}. These curves are shown in Figure \ref{fig2}. This first analysis was performed regardless the context and sentimental connotation in which the phrases appeared.

We calculate the Spearman's correlation coefficient \cite{lehman2005jmp} between the total number of mentions of each candidate (Figure \ref{fig2}) with the spread between polls data (Figure  \ref{fig1} (b)). We take into account that changes in media coverage may not be instantaneously reflected in the polls, either because 
the time scale of how media can exert influence is not clear, or the publication date of the polls is posterior to the data collection. Therefore, we calculate a lagged correlation between the time series for a range of lags.

\par We found that the number of mentions of both candidates in New York Times, CNN, and USA Today correlates positively with the spread between Clinton and Trump in the polls, with an average correlation coefficient of $0.663$ (NYT), $0.426$ (CNN), and $0.246$ (USA) respectively. This means that when the number of mentions increases in  these outlets, the difference Clinton minus Trump increases, no matter which candidate is mentioned.
On the other hand, these correlations are negative for similar time series in Fox News, with an average correlation coefficient of $-0.476$.
In this case, when the number of mentions of any of the candidates increases in Fox News, the difference Clinton minus Trump decreases. 
These results are statistically significant ($p < 0.001$) for a lag between $7$ and $15$ days for The New York Times, Fox News and CNN, while for USA Today these are significant for a lag between $12$ and $15$ days ($p < 0.05$).
Similar conclusions can be achieved when studying the mentions of the candidates separately.

\par The correlation signs depend on the media independently of candidates. In order to go deeply in the causes of this behavior, we apply sentiment analysis and topic modeling on the articles.

\subsection{Sentiment analysis}
\label{sec:sentiment_analysis}

To study the connotation with which each candidate is mentioned, we applied the sentiment classifier algorithm described above. The procedure is the following:

\begin{enumerate}
\item In each text, we detected phrases mentioning terms ``Hillary", ``Clinton", ``Donald" or ``Trump". In the case when more than one candidate is mentioned, we separated the sentences using syntactic analysis.
\item We applied the sentiment analysis for these sentences and counted the amount of positive, negative, and neutral mentions for each of the candidates. In this step, the deep recursive models for the semantic composition play a central role, since the syntactic analysis of a sentence allows  to understand when the text refers to a given candidate in a positive or negative way.
\end{enumerate}

\par After this procedure, we registered for every day in the studied period, the number of phrases related to each candidate, as well as their sentiment score. Based on this classification, we define a Sentiment Bias statistic $SB$ (Eq. (\ref{eq:SB})), where $\#C_+$ ($\#C_-$) stands for fraction of positive (negative) mentions of Hillary Clinton and $\#T_+$ ($\#T_+$) for positive (negative) mentions of Donald Trump in a given mass media outlet. 
$SB$ is a measure of the  bias towards one of the candidates: If $SB>0$, the bias is positive towards Clinton compared with Trump, and on the other hand, if $SB<0$, the bias is positive towards Trump.

\begin{eqnarray}
\label{eq:SB}
    SB = (\#C_+ - \#C_-) - (\#T_+ - \#T_-)
\end{eqnarray}

We calculated the value of $SB$ for each media. 
The results of this analysis are $SB_{NYT} = 0.162 \pm 0.004$ for The New York Times, $SB_{USA} = 0.160 \pm 0.010 $ for USA Today, $SB_{CNN} = 0.094 \pm 0.006$ for CNN, and $SB_{FN} = 0.046 \pm 0.005 $ for Fox News. In all cases, we reject that $SB$ is a negative value ($p < 0.001$, by bootstrapping  \cite{efron1994introduction,efron2003second}, see appendix \ref{sec:bootstrapping}). 
\par Although in all cases $SB$ is a positive value, the Sentiment Bias statistic is significantly low for Fox News, while we did not find significant differences between The New York Times and USA Today. 
In summary, we found that $SB_{NYT},\:SB_{USA} > SB_{CNN} > SB_{FN}$. 
For instance, this suggests that The New York Times, USA Today, and CNN seem to mention Hillary Clinton in a more positive way than Fox News.
\par The fact that $SB$ is positive for all media suggest that this analysis alone is not enough to understand  the behavior of the  spread between Clinton and Trump in the polls. The temporal dependence of the Sentiment Bias statistic ($SB = SB(t)$) will provide a useful insight to understand differences in behavior between both media. In order to  go further in this direction we need to analyze first  the covered topics and their relative relevance.

\subsection{Topic analysis}

\par We classified the news corpus in six topics and called Media Agenda the relative importance that each media outlet gives to the set of topics, as calculated by Eq. (\ref{eq:topic_weight}) and defined in \cite{Pinto2019}.

\par  The first topic is about elections in general and it is represented by words like campaign, election, candidate, etc. This result is consistent with the fact that we analyze political news during the campaign period. We choose to discard it given its lack of specificity. 
The other five topics reveal the subjects discussed during the electoral campaign, which we label and describe briefly as: 
\begin{itemize}
    \item ``Clinton email controversy": covers the famous controversy which Hillary Clinton faced during the elections due to the use of her private email server for official communication.
    \item ``Clinton Foundation affair": is about the allegations of possible conflicts of interest due to the fact that Clinton was Secretary of State and her foundation’s accepted foreign donations.
    \item ``Economy": discusses  particularly on taxes, income, jobs and business.
    \item ``Immigration": is about the discussion of immigration policies between Mexico and USA, raised in the Donald Trump's campaign.
    \item ``Foreign affairs": deals with the United States foreign policy. In particular, it centers on ISIS and the hypothetical interference of Russians in the electoral process.
\end{itemize}

\par
\begin{figure}[h]
\centering
\includegraphics[width = \textwidth]{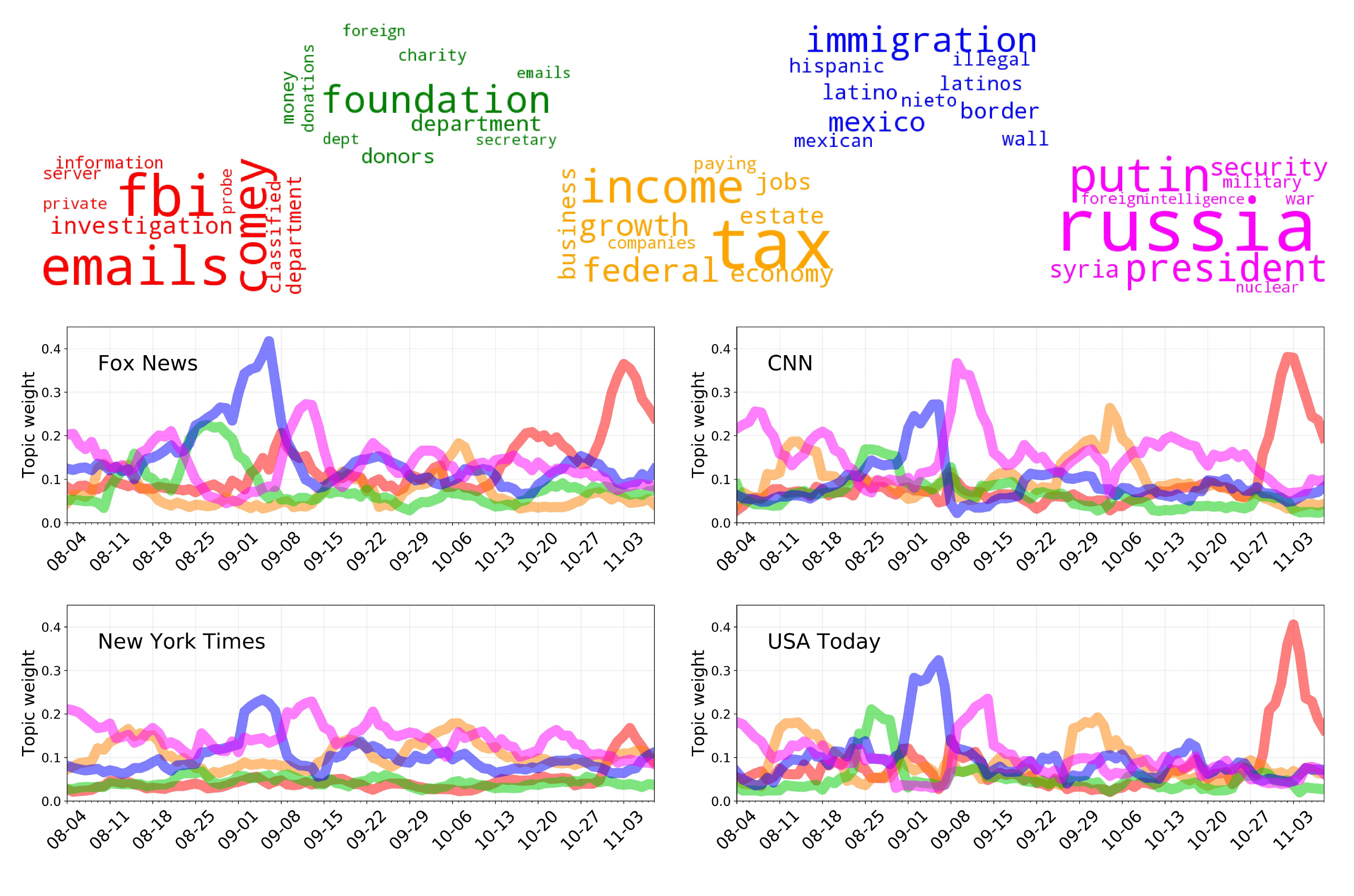}
\caption{\textbf{Topic coverage for each media outlet.}
 We show the time evolution of the topics during the whole period, where the difference between these two media outlets can be also observed. Each topic is specified by the word-clouds on top of the figure (and also pointed out in the main text). Topic 1 (red): Clinton email controversy; Topic 2 (green): Clinton Foundation affair; Topic 3 (yellow): Economy; Topic 4 (blue): Immigration; Topic 5 (magenta): Foreign Affairs.}
\label{fig3}
\end{figure}

 The keywords which define these five topics are represented in the word clouds of the top panels in Figure \ref{fig3} and correspond to the most significant words which describe the similarity among the news articles grouped in a given topic. It is worth noting that this is an unsupervised method and therefore keywords emerge from  the analyzed corpus of news and were not arbitrarily chosen. Figure \ref{fig3} also shows the temporal evolution of these topics. 

\par
\begin{figure}[h]
\centering
\includegraphics[width = \textwidth]{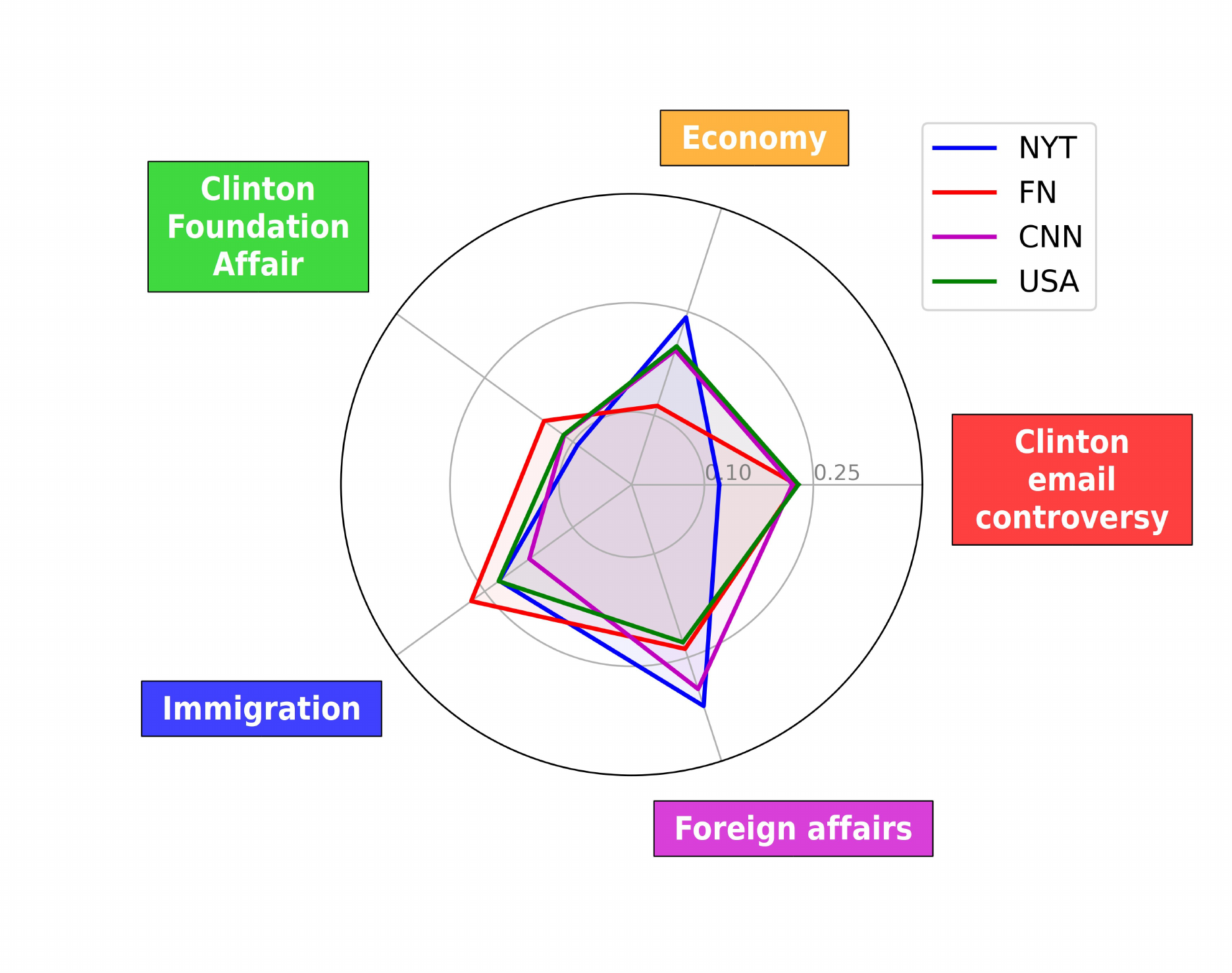}
\caption{\textbf{Topic coverage for each media outlet over the whole period.} 
The radar plot shows the accumulative coverage, i.e. the agenda of each media outlet over the five topics.}
\label{fig4}
\end{figure}

 The comparative Agenda of each media can be easily visualized in Figure \ref{fig4}. We observe that the New York Times emphasizes the topics \emph{Economy} and \emph{Foreign Affairs}, while Fox News gives more coverage to \emph{Clinton's affairs} and \emph{Immigration}. 
On the other hand, we can see that both CNN and USA Today cover \emph{Economy}, sharing this interest with New York Times, but also pay attention to the topic \emph{Clinton email controversy} as Fox news does.

\par As we did before with the number of mentions, we calculate the Spearman's correlation  for a range of lags between the topic's coverage evolution and the spread between candidates. 
We found almost all  correlation coefficients are significant for lags around 10 or 15 days, except Foreign Affairs which does not show significant correlation for three of the four media outlets considered (see Table \ref{table:correlations}).

\begin{table}[!ht]
\centering
\caption{\textbf{Linear correlation}.}
\begin{tabular}{|l|c|c|c|c|}
\hline
Topic & NYT (SRL) & Fox (SRL) & CNN (SRL) & USA (SRL)\\ \hline
\hline
Clinton email controversy & -0.46 (10-20) & -0.42 (11-20) & -0.45 (13-20) & -0.54 (10-20)\\ \hline
Economy & 0.56 (4-20) & 0.59 (8-20) & 0.48 (5-18) & 0.40 (10-15) \\ \hline
Clinton Foundation affair & -0.53 (3-20) & -0.43 (15-20) & -0.53 (1-20) & -0.40 (5-12)\\ \hline
Immigration & -0.42 (0-12) & -0.44 (5-20) & - & -0.47 (12-20)\\ \hline
Foreign affairs & 0.44 (17-20) & - & - & - \\ \hline
\end{tabular}
\begin{flushleft} Average Spearman's correlation coefficient between topic coverage and spread between candidates on statistically significant range of lags (SRL) ($p < 0.001$). Non-significant values are not reported.
\end{flushleft}
\label{table:correlations}
\end{table}

\par The results shown in Table \ref{table:correlations} indicate that the context in which the candidates are mentioned plays a key role in the correlation with the intention to vote. For instance, the time series corresponding to topics \emph{Email controversy}, \emph{Foundation affair} and \emph{Immigration} negatively correlates with the spread between candidates (i.e. the coverage of these three topics worsens Clinton's image), conversely to what happens with the topics \emph{Economy} and \emph{Foreign affairs} in New York Times. 
Notoriously, as we can see in Figure \ref{fig4}, these two topics are more emphasized by New York Times and CNN, and the first one also by USA Today, while Fox News covers with more intensity the former three, which appears to affect Clinton's intention.
These results are consistent with the calculation of $SB$ in the previous section.

\subsection{Combining Sentiment and topic analysis}
\label{sec:TopicSentiment}

\par
\begin{figure}[h]
\centering
\includegraphics[width = \textwidth]{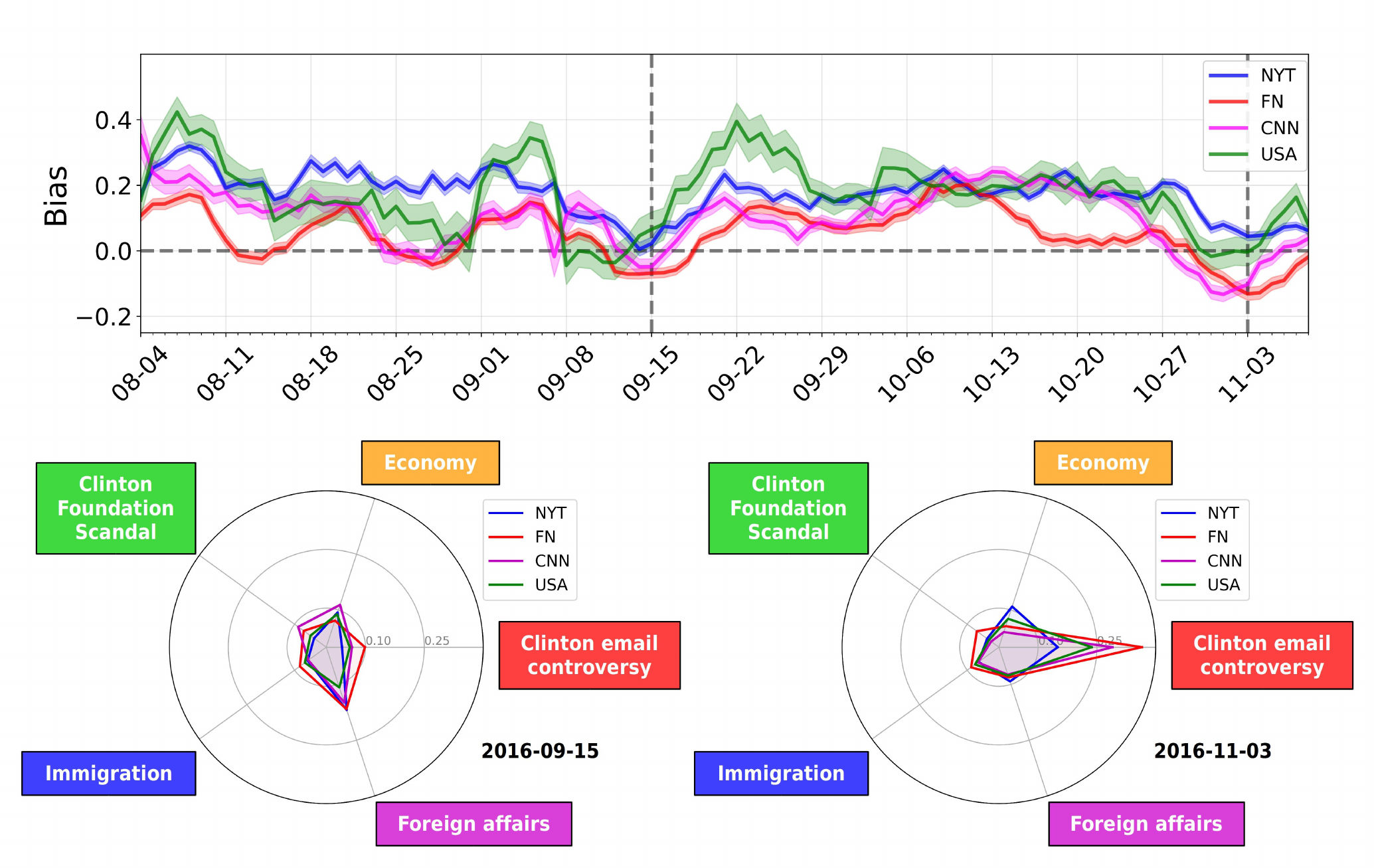}
\caption{\textbf{Time series of sentiment bias and radar plots}.
The curves on top shows time evolution of the Sentiment Bias statistic $SB$ for each newspaper, New York Times in blue, Fox News in red, CNN in magenta, and USA Today in green. The figures on the bottom are the radar plots of the coverage over the five topics for two specific days: 2016-09-15 and 2016-11-03 (vertical dashed lines in top panel), which correspond to the largest negative values of SB for Fox News and CNN.}
\label{fig5}
\end{figure}

 Although average Sentiment Bias statistic (SB) is positive regardless of the media outlet, there exists a period of time where for instance, $SB_{FN}(t)<0$ or $SB_{CNN} < 0$ (See Figure \ref{fig5}).  Here we propose a combined analysis in order to better understand this behavior.

\par In Table \ref{table:SBtopic}, we calculated the Sentiment Bias statistic (SB) for each topic, discriminated for media outlet. We  observe that in the first three topics the sign of $SB$ matches the sign of the correlation displayed in Table \ref{table:correlations}, while the sign of the last topic matches with the only significant correlation of that table. The only mismatching is in the topic \emph{Inmigration}.
These results reveal that sentiment analysis is much more informative when news articles are decomposed in topics than grouped all together.

\begin{table}[!ht]
\centering
\caption{\textbf{The sentiment bias per topic.}}
\begin{tabular}{|l|c|c|c|c|}
\hline
Topic & $SB_{NYT}$ & $SB_{FN}$ & $SB_{CNN}$ & $SB_{USA}$\\  \hline
\hline
Clinton email controversy & -0.475 & -0.429 & -0.302 & -0.315 \\ \hline
Economy & 0.332 & 0.070 & 0.152 & 0.168 \\ \hline
Clinton Foundation Scandal & -0.256 & -0.257 & -0.304 & -\\ \hline
Immigration & 0.501  & 0.347 & 0.306 & 0.382 \\ \hline
Foreign affairs & 0.146 & 0.053 & 0.115 & 0.166\\ \hline
\end{tabular}
\begin{flushleft} The Sentiment Bias statistic $SB$ calculated with the news of each topic and each newspaper together with the sign of the correlation of the same topic with the difference between Clinton and Trump in the polls. In all cases we reject the hypothesis that $SB$ has an opposite sign with $p < 0.001$. Non-significant values are not reported.
\end{flushleft}
\label{table:SBtopic}
\end{table}

\par The success of this combined analysis can be seen again in Figure \ref{fig5} where, in addition to the time evolution of $SB$ for each outlet, we can see the radar plots of the agendas for two specifics dates, which belong to periods where  $SB_{FN}(t),\: SB_{CNN} <0$.
In those dates, we can see that the difference between agendas can be partially explained by the topic Clinton emails controversy, which was more emphasized by Fox News and CNN than New York Times, more notorious at the end of the period.

\subsection{Causality}

\par In this section, we look for a causal relationship between the spread of polls  ($CT(t)$) and the time series of the topics by applying the Granger causality framework described in section \ref{sec:Granger}.
Due to the fact the $G$ is not a stationary series, we start by calculating its first difference $\Delta CT$, which is stationary (Augmented Dickey-Fuller test \cite{seabold2010statsmodels}, $p < 0.05$) and therefore can be modeled by autoregressive models. 
By studying the full and partial auto-correlation of $\Delta CT$, we noticed that it is essentially described by a random-walk according to Eq. (\ref{eq:gap}), with $w_t$  a standard normally distributed random value.
We propose two models for the causal analysis: One describe by Eq. (\ref{eq:gap}), and the other including the information about topics coverage within a certain lag $\tau$ (Eq. (\ref{eq:gap_topics}), where $T_i (t)$ is the weight of topic $i$ at time $t$). 

\begin{eqnarray}
\label{eq:gap}
\Delta CT(t) = CT(t) - CT(t-1) = w_t
\end{eqnarray}

\begin{eqnarray}
\label{eq:gap_topics}
\Delta CT(t + \tau) = \beta \cdot \Delta T_i(t) + w_{t + \tau}
\end{eqnarray}

\par We say that the coverage of a given topic effectively affects the spread between candidates when the parameter $\beta$ in Eq. (\ref{eq:gap_topics}) differs significantly from zero. Since both models involved only first differences, the proper interpretation is that a non-zero value of  $\beta$ implies that the growth or decrease of a given topic predicts a variation in the spread after a certain lag. 

\par  The topics with $\beta$ significantly different from zero ($p < 0.01$), 
with a sign and lag consistent with the results of linear correlations calculated in previous sections,
are: \emph{Clinton's email controversy} ($\beta < 0$ and $\tau = 19$) for Fox News;
\emph{Economy} ($\beta > 0$ and $\tau$ between $11$ and $16$) for Fox News, CNN, and USA Today;
and \emph{Clinton Foundation affair} ($\beta < 0$ and  $\tau = 19$) and
\emph{Inmigration} ($\beta < 0$ and $\tau = 10$), both for The New York Times.

\par Finally we would like to remark the role of the topic \emph{Clinton's email controversy}.
The analyzed data suggests that this topic plays a key role in the period of time close to the election day. This can be observed in the Fox News and CNN coverage of Figure \ref{fig3} and in the radar plots of Figure \ref{fig5}, where  this topic had a larger coverage during the last week. 
Our analysis suggests that, when this topic becomes the most important in news outlets, there is a notorious reduction  in the difference between Clinton and Trump. This is the reason why our model reports a causal relationship between this topic and the difference of the surveys. Interestingly, that happens at the end of the period, which suggest that this was a key topic in the electoral result.

\section{Conclusions}
\label{sec:discussion}

\par The influence of mass media on public opinion has been studied from different perspectives and methodologies, ranging from field experiments to data analysis and computational models. 
The vast availability of data coming from mass media communication and social media  makes the analysis techniques based on data be important in the investigation of this kind of issues. 

\par In this paper we suggest a set of tools grounded on natural language processing techniques to be applied in the analysis of the effects that mass media can produce on public opinion. In particular, we are interested in addressing which features of news articles are related to measurable changes in public opinion, using sentiment analysis and topic detection of news articles. Specifically,  either the sentiment content of the news articles or the topic of the mentions is important. 

\par Each method on its own, sentiment analysis and topic decomposition, has been widely applied to study related problems. However, to our knowledge, the use of the combination of the two methods to analyze the impact of mass media has not been done yet in systematic way.  

\par By applying the developed methodology to the media coverage of the 2016 US presidential elections,  we can understand key aspects in the relationship between mass media and public opinion. In this example, we analyzed news articles in which at least one of the two candidates involved were mentioned. We performed a sentiment analysis and topic detection on that corpus and compared them with measures of vote intention as a proxy of public opinion.

\par Our approach allows to extract useful insights to this example, as can be seen in the list bellow:

\begin{itemize}
    \item The total number of mentions of both candidates in news articles correlates positively with the difference between Clinton and Trump in the polls in the New York Times, CNN, and USA Today, but negatively in Fox News, independently of the candidate.
    \item The average sentiment analysis of the news articles where both candidates were mentioned is not enough to explain previous behavior.
    \item The topic analysis, which allows us to appreciate the difference between the agenda of both media outlets, shows that the coverage of given topics are correlated with the difference between Clinton and Trump in the polls.
    \item The sentiment analysis discriminated by topic is consistent with these last results (except for Immigration), given that the topics in favor of Clinton shows positive sentiment towards Clinton and those in favor of Trump are negative towards Clinton (or positive towards Trump).
    \item There is a causal relation ($p<0.05$) in the Granger sense between four topics and the difference between Clinton and Trump in the polls, that means that these topics serve as good predictors for the variations in the polls.
    \item The topic related to the Clinton email controversy seems to be the most relevant because:
    \begin{enumerate}
        \item It negative correlates with the difference between Clinton and Trump in the polls.
        \item It has a significant causal relation with the difference between Clinton and Trump in the polls, which, as we said before, determines this topic as a good predictor for the variation in the polls.
        \item It explains the negative values of the Sentiment Bias statistic of Fox News ($SB_{FN}(t)$) and CNN ($SB_{CNN}(t)$).
    \end{enumerate}
\end{itemize}

\par It is important to notice that the choice of media outlets does not limit the analysis performed in this manuscript, given than we can add as many media as necessary. However, it could be interesting to analyze the role of social networks in the relation between news consumption and public opinion.
Finally, it should be mentioned that  the methodology implemented in this work is far general and  can be easily extended to other topics of interest, not necessarily restricted to a political scenario.

\section{Data Availability}
\indent
The data that support the findings of this study are available upon request from the authors.

\section{Acknowledgments}
We thank Marcos Trevisan for careful reading of the manuscript and helpful comments. This work was partially funded by the Agencia Nacional de Promoción Científica of Argentina via grant PICT 201-0215.

\section{Figures}

\clearpage

\bibliographystyle{unsrt}

\clearpage

\appendix
\renewcommand\thefigure{\thesection.\arabic{figure}} 
\section{Mass Media Trends}
\label{sec:Finite_size_scaling}
\setcounter{figure}{0} 

\par
\begin{figure}[h]
\centering
\includegraphics[width = \textwidth]{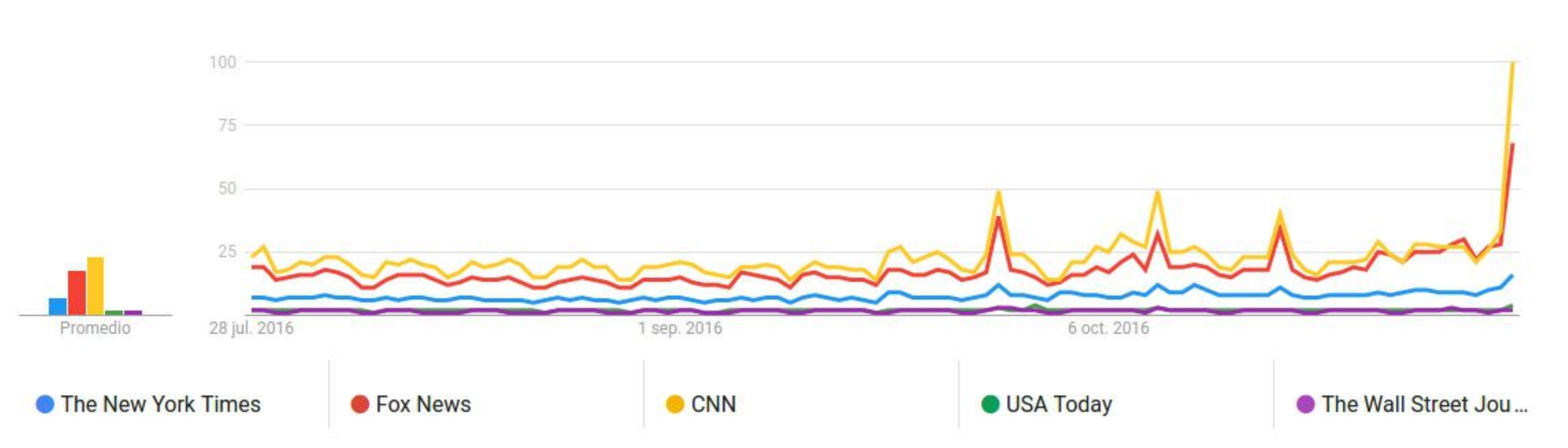}
\caption{\textbf{Google Trends of the Mass Medias}. 
Left: The mean relative frequency of searches of five Mass Media between the $28^{th}$ of July and the $8^{th}$ of November; Right: The corresponding time series of the relative frequency of searches. The Google trends search is limited to the United States.}
\label{figA1}
\end{figure}

 In this work we analyzed the role of the Mass Media and its influence in society during elections. Consequently, we selected Mass Media that are highly consumed by the American population. Therefore, Google Trends, an official Google tool that compares the frequency with which different terms are searched, was used. In particular, we compared newspapers and news portals of the United States. The list includes: ABC, Boston Globe, CBS, Chicago Tribune, CNN, Fox News, Houston Chronicle, Los Angeles Times, New York Daily News, New York Post, Newsday, NPR, Tampa Bay Times, The Dallas Morning News, The Denver Post, The New York Times, The Wall Street Journal, The Seattle Times, USA Today and Washington Post. Among the two most important were Fox News and New York Times, as can be seen in Figure \ref{figA1}. The search in google trends was filtered by geographic location, limiting it only to the United States, and by period of time, from the $28^{th}$ of July (2016 Democratic National Convention) until the $8^{th}$ of November (the day of the election).

\clearpage

\section{Bootstrapping}
\label{sec:bootstrapping}

\par To test the significance of the Sentiment Bias ($SB$) calculated in section \ref{sec:sentiment_analysis}, as well its error bars, we employed the bootstrapping technique \cite{efron1994introduction,efron2003second}).
It consists on approximating the unknown probability distribution of a given statistic by sampling with replacement from the data. 
By this way, one can construct confidence intervals in order to test the significance of a given result.
\par For instance, if we have the following data: 3 positive mentions and 1 negative mention for candidate A, while 2 positive mentions for candidate B and 3 negative mentions,
this can be represented as [A+, A+, A+, A-, B+, B+, B-, B-, B-] (where A+ means a positive mention of candidate A).
In this case, we would obtain $SB =  1/3$ (taking $SB$ positive as favoring candidate A, see Eq. (\ref{eq:SB})).
Then, we generate new data sets by sampling with replacement from the original one, as many elements as its length.
For example, a generated data set could be [A+, A+, A+, A+, A+, B-, B-, B+, B+], where $SB = 5/9$.
By successive repetitions of this process, we finally obtain an approximate distribution for $SB$.

\end{document}